\documentclass[twocolumn,showpacs,amsmath,amssymb,pre]{revtex4}

\usepackage{graphicx}
\usepackage[dvipsnames]{xcolor}
\usepackage{subfigure}
\usepackage{txfonts}
\usepackage{dcolumn}
\usepackage{bm}
\usepackage[latin1]{inputenc}  
\usepackage[T1]{fontenc}       

\begin{document}

\title{Anisotropic nonlinear elasticity in a spherical bead pack:
influence of the fabric anisotropy}

\author{Yacine Khidas}
 \email{yacine.khidas@univ-mlv.fr}
\author{Xiaoping Jia}
 \email{jia@univ-mlv.fr}
\affiliation{Université Paris-Est, Laboratoire de Physique des
Matériaux Divisés
 et des Interfaces, CNRS UMR 8108, Cité Descartes, 77457 Marne-la-Vallée, France }


\begin{abstract}
Stress-strain measurements and ultrasound propagation experiments in
glass bead packs have been simultaneously conducted to characterize
the stress-induced anisotropy under uniaxial loading. These
measurements, realized respectively with finite and incremental
deformations of the granular assembly, are analyzed within the
framework of the effective medium theory based on the Hertz-Mindlin
contact theory. Our work shows that both compressional and shear
wave velocities and consequently the incremental elastic moduli
agree fairly well with the effective medium model by Johnson
\emph{et al.} [J. Appl. Mech. \textbf{65}, 380 (1998)], but the
anisotropic stress ratio resulting from finite deformation does not
at all. As indicated by numerical simulations, the discrepancy may
arise from the fact that the model doesn't properly allow the grains
to relax from the affine motion approximation. Here we find that the
interaction nature at the grain contact could also play a crucial
role for the relevant prediction by the model; indeed, such
discrepancy can be significantly reduced if the frictional
resistance between grains is removed. Another main experimental
finding is the influence of the inherent anisotropy of granular
packs, realized by different protocols of the sample preparation.
Our results reveal that compressional waves are more sensitive to
the stress-induced anisotropy, whereas the shear waves are more
sensitive to the fabric anisotropy, not being accounted in
analytical effective medium models.
\end{abstract}


\maketitle

\section{\label{sec:intro}Introduction}
The study of nonlinear elasticity in granular materials is not only
of fundamental interest but also of practical importance in many
fields, including soil mechanics and geophysics. Unlike consolidated
porous materials such as sedimentary rocks, soils and sintered bead
packs, dry granular materials acquire solely its elasticity as a
result of the applied stress, forming a very inhomogeneous force
network \cite{goddard90, jaeger96b, degennes99, makse04,
goldenberg05, jiang07}. The mechanical properties of such materials
depend strongly on both the contact interactions between grains and
the geometric arrangement (i.e. packing structure). The stress
dependence of elasticity which originates with these contact forces
at the grain level leads to a complex behaviour of the granular
medium: nonlinear behaviour, loading-path dependence and
stress-induced anisotropy \cite{norris97}. The change of sound
velocity with the applied stress (i.e. acousto-elastic effect) is
the characteristic signature of this nonlinear elasticity
\cite{goddard90,makse04,jia99,somfai05}. The hysteresis often found
in the stress-strain experiments of granular media is also
associated with the path-dependant nature of the contact forces
\cite{jiang07,hoque98,garcia05}. Another issue complicating the
analysis of the mechanical properties of granular materials arises
from the geometric or fabric anisotropy, which is basically related
to the mode of grain deposition and the packing structure in a model
system like the sphere pack \cite{hoque98,hicher06}. This fabric
anisotropy is important for the mechanical behaviour of granular
media such as stress transmission \cite{vanel99}, mechanical
stability \cite{grasselli97} and liquefaction resistance
\cite{ishibashi03}.

Photoelastic visualizations \cite{majmudar05} and numerical
simulations in 2D systems
\cite{radjai01,cambou04,luding04,peyneau08} demonstrate that the
anisotropy induced by an external load may have two distinct
effects: one leads to an important change of the fabric anisotropy
in the contact network, and the other develops an anisotropic force
chain network. Understanding the nonlinear elastic responses and the
associated stress-induced anisotropy represents a fundamental issue
for granular mechanics. The problem may also be important for
understanding the jamming phenomena and shear-induced yielding in a
more general class of geomaterials formed from granular media
\cite{grasselli97,ishibashi03,majmudar05}. Development of non
destructive methods of investigation is therefore desirable for
monitoring the evolution of both geometric and mechanical
anisotropies in real 3D granular materials, and allows us to gain a
more comprehensive insight into their elastic properties along
different stress paths.

Sound waves offer a sensitive and non invasive probe of both the
structure and the mechanical properties of heterogeneous materials
\cite{toksoz81,winkler83,guyer99,gilcrist07}. In a granular medium,
sound propagation is controlled to a large extent by the properties
of the contact force networks forming the solid frame of the
material \cite{makse04,jia99,liu92,jia04,bonneau07}. The effective
medium theory (EMT) has been commonly used to describe both the
nonlinear elasticity
\cite{goddard90,jiang07,hicher06,walton87,johnson98} and sound
propagation in the \emph{long-wavelength} limit
\cite{goddard90,makse04,duffy57,digby81}. These effective medium
models are generally based on the Hertz-Mindlin theory at the grain
contact and make use of the two main assumptions to obtain a mean
field description of the granular elasticity: i) affine
approximation or kinematic hypothesis, in which the motion of each
grain follows the applied strain, and ii) statistically isotropic
distribution of contacts around each grain. However, some acoustic
measurements in random packs of glass beads under isotropic loading
\cite{makse04,domenico77} show that the ratio of bulk modulus $K$ to
shear one $G$ is significantly larger than the value predicted by
the EMT with bonded elastic spheres (no sliding). Numerical
simulations reveal that the discrepancies can stem from the failure
of the effective medium approximation: the bulk modulus is well
described by the EMT but the shear modulus is not, principally
because the EMT does not correctly allow the grains to relax
collectively from the affine motion assumed by the theory
\cite{makse04,jenkins04}. In a recent work, the multiple scattering
of \emph{shear} acoustic waves through the stressed glass bead packs
shows that sound propagation at \emph{small-amplitude} (or
incremental deformation) does not cause any significant
rearrangement of the contact force network \cite{jia04}. This
observation leads to a fundamental question as to whether
overestimating the shear modulus by the EMT is related with the
failure of the effective medium approximation for sound propagation
or with the inadequate treatment of micromechanics at the grain
contact level \cite{norris97,winkler83}.

In this paper, we present new results obtained from a granular model
system, i.e. a glass bead pack in an {\oe}dometric test. We
conducted simultaneously the stress-strain measurement and the
ultrasound propagation experiment to characterize the stress-induced
anisotropy by uniaxial loading. Both measurements of stress fields
and elastic moduli via sound velocities parallel and perpendicular
to the applied load are analyzed within the framework of the EMT
developed by Johnson \emph{et al.} specifically for stress induced
anisotropy in the {\oe}dometric configuration \cite{johnson98}. The
aim of our work is twofold. Firstly, we examine the applicability of
the EMT to small- and large-deformation mechanical tests,
corresponding respectively to sound propagation and stress-strain
measurements. The crucial role of the interaction nature at the
grain contact for the appropriate prediction will be discussed.
Then, we investigate the respective responses of compressional and
shear waves to the elastic anisotropy in granular media. A
particular attention is paid to the influence of the fabric
anisotropy which is realized by the different sample preparation.
Such inherent fabric anisotropy is expected to evolve little under
{\oe}dometric loading compared to pure shear experiments.
\section{\label{sec:experiment}Experiments}
\begin{figure}[]
  \includegraphics[width=8cm]{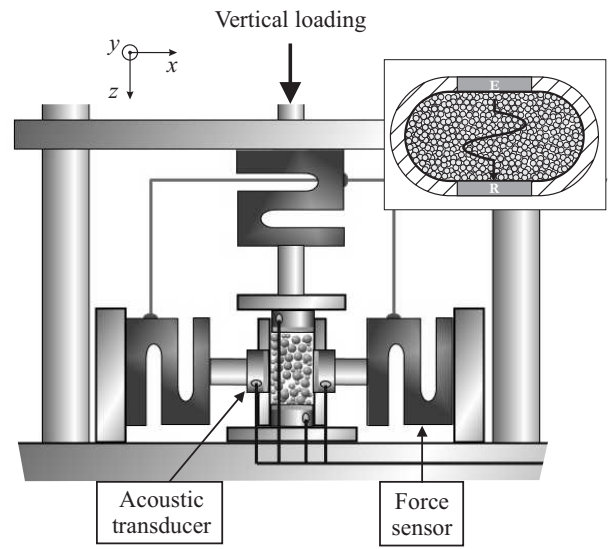}
  \caption{\label{fig:apparatus}Sketch of the experimental set-up.
   Inset: Top-view of the {\oe}dometric cell.}
\end{figure}

In order to characterize the anisotropic elasticity of granular
materials, we developed an apparatus coupling the mechanical test
with the ultrasonic measurement. The schematic diagram of the
experimental arrangement is illustrated in Fig. \ref{fig:apparatus}.
Spherical glass beads of diameter between 300 $\mu$m and 400 $\mu$m
are carefully filled in a duralumin cell up to 70 mm in height and
closed by the fitted pistons at the top and bottom surfaces. A
controlled vertical load is applied to the granular sample across
the top piston. The forces re-directed in the horizontal direction
are measured by the stiff force sensors placed at the middle of the
cell to monitor the stress anisotropy evolution. This special
compaction cell is presumed to be rigid enough that the sample
experiences a macroscopic uniaxial strain with negligible lateral
motion ({\oe}dometric test). The solid volume fraction $\phi$ of the
granular assembly is determined by the axial displacement as a
function of the applied load. To ensure the measurements of sound
velocities in both the vertical ($z$-axis) and horizontal ($x$-axis)
directions, a pair of compressional or shear piezoelectric
transducers of diameter 30 mm in contact with the glass beads are
placed at the top and bottom pistons, while the other ones are
mounted at the lateral walls separated by a distance of 40 mm (see
the inset of Fig. \ref{fig:apparatus}).
\subsection{Granular fabric and packing density }
To study the influence of the granular fabric on the induced elastic
anisotropy, the bead packs are realized by two different preparation
protocols. The first one, known as rain deposition or air
pluviation, consists in pouring the glass beads into the cell
through two grids of mesh size 2 mm spaced by 2 cm and reaches a
dense pack of $\phi \approx 0.64$. The second, termed as
de-compaction protocol is to remove gently a horizontal grid through
the bead pack from the bottom to the top after filling the beads in
the cell; the rearrangements produced by this efficient shearing
throughout the sample allow us to obtain a loose pack of $\phi
\approx 0.60$. It is shown by numerical simulations
\cite{radjai01,emam05} that the rain deposition protocol creates an
anisotropic distribution of the contact angle with two preferred
directions orientated roughly at 30° around the vertical (gravity)
direction. However, the loose pack prepared by the de-compaction
protocol is expected to produce a fairly isotropic distribution of
the contact angle \cite{reydellet02}.

\begin{figure}
  \includegraphics[width=8cm]{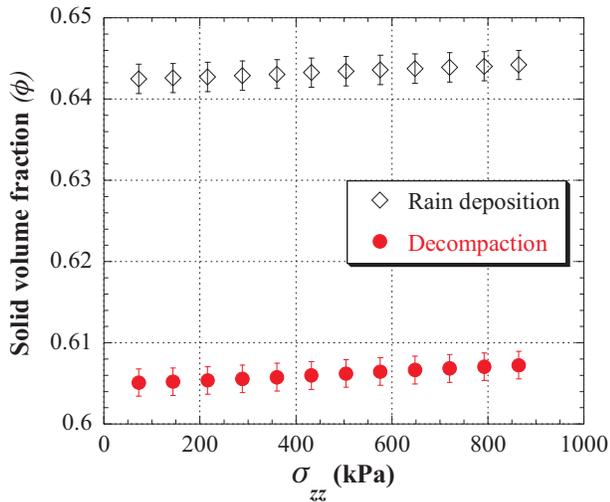}
  \caption{\label{fig:compac}Evolution of solid volume fraction of the loose and dense packs.
   Error bars illustrate the data dispersion over 30 measurements.}
\end{figure}

Before any measurement, a preloading up to $\sigma_{zz} = 400$ kPa
is applied to the sample in order to minimize the hysteretic effects
related to the grain rearrangements and ensure a reproducible
initial state. Then the stress-strain and ultrasonic velocity
measurements are performed as a function of the applied stress
$\sigma_{zz}$ ranging from 70 to 900 kPa. Figure~\ref{fig:compac}
displays the evolution of the solid volume fraction $\phi$ versus
$\sigma_{zz}$ in the dense and loose packs, obtained respectively by
the two preparation protocols. These results show a good
reproducibility of our sample preparation in terms of the initial
packing density measured at $\sigma_{zz} = 70$ kPa for 30 repeated
measurements: $\phi_{dense} = 0.642 \pm 0.002$ and $\phi_{loose} =
0.605 \pm 0.002$. The fact that the solid volume fraction in the
dense packing sample is a little larger than the density of the
random close packing (RCP) is due to the slight dispersion of the
bead size. Furthermore, Fig.~\ref{fig:compac} shows that the packing
density of our samples increases slightly, up to a resultant
vertical deformation of $\varepsilon_{zz} (\equiv \varepsilon_{3}) =
3\times10^{-3}$ at the applied $\sigma_{zz} (\equiv \sigma_{3}) =
900$ kPa. Both the dense and loose bead packs keep clearly a
signature of their preparation protocols. These results imply that
for the applied stress in this work, much less than the value to
produce the grain fracture of about 20 MPa for the glass beads
\cite{mcdowell02}, there are few important rearrangements of grains
in the {\oe}dometric test as compared to those in a pure shear
experiment \cite{majmudar05} and the inherent fabric should not
change much from the initial state \cite{cambou04}.
\subsection{Anisotropic stress field}
\begin{figure}[]
  \includegraphics[width=8cm]{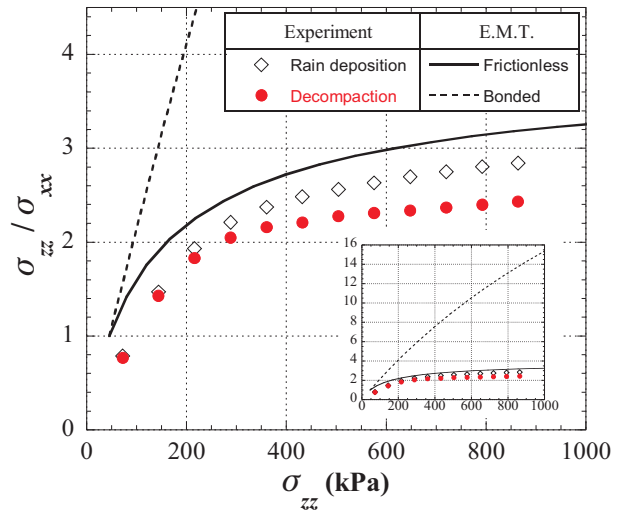}
  \caption{\label{fig:StressRatio}Evolution of the vertical-to-horizontal stress ratio $\sigma_{zz} / \sigma_{xx}$
  versus the applied stress $\sigma_{zz}$ in the dense and loose packs.}
\end{figure}

Let us now investigate the stress-induced anisotropy under uniaxial
loading in the {\oe}dometric configuration. By monitoring the
resultant horizontal stress $\sigma_{xx}$ when increasing the
vertical stress $\sigma_{zz}$, we display in Fig.
\ref{fig:StressRatio} the stress ratio $\sigma_{zz} / \sigma_{xx}$
measured as a function of the applied stress $\sigma_{zz}$ for the
two granular samples, respectively. It is observed that both the
bead packs evolve from their initial stress fields at low applied
$\sigma_{zz} = 70$ kPa, roughly isotropic, to anisotropic states at
high  $\sigma_{zz}$. As mentioned in the previous experiment
\cite{johnson98}, the isotropic distribution of stress observed at
low $\sigma_{zz}$ arises probably from a kind of tight wedging of
grains produced during the preloading and unloading cycle. Such an
interlock of grains may contribute to a residual stress
isotropically distributed in the bead pack even when the top piston
is removed and no stress is applied.

As the applied stress increases, the ratio $\sigma_{zz} /
\sigma_{xx}$ is seen to increase to an asymptotic value at high
$\sigma_{zz}$. For the granular packs prepared by the two distinct
protocols, the different asymptotic values of about $11\%$ reveal
again a memory effect of the initial state of the sample as seen in
the above density measurement. In the conventional {\oe}dometric
test, the inverse of the asymptotic value $\sigma_{zz} /
\sigma_{xx}$ is known as Jacky coefficient $K_0$ of the earth
pressure at rest; it is empirically related to the internal friction
angle $\theta_0$ of the medium by $K_0 = 1 - \sin \theta_0$
\cite{michalowski05}.  These results thus suggest a possible
correlation between the internal friction angle and the fabric of
the medium. To examine this issue, we have measured the avalanche
angle $\theta$ at the relative humidity $RH = 40\%$ for both the
dense and loose packs, and obtained  $\theta_{loose}=30.4 \pm 1.0°$
and  $\theta_{dense}=33.2 \pm 0.3°$ for six avalanche experiments.
The difference between the two measured avalanche angles is about
$9\%$, consistent with the difference of $11\%$ between the two
internal friction angles deduced from the asymptotic values of
$\sigma_{zz} / \sigma_{xx}$ in Fig. \ref{fig:StressRatio}. Although
the avalanche angle is not the direct measure of the internal
friction angle, these stress measurements indicates again the
significant effect of the preparation protocol on the granular
fabric \cite{vanel99,evesque93}.
\subsection{Anisotropic elastic moduli}
\begin{figure}
  \includegraphics[width=7cm]{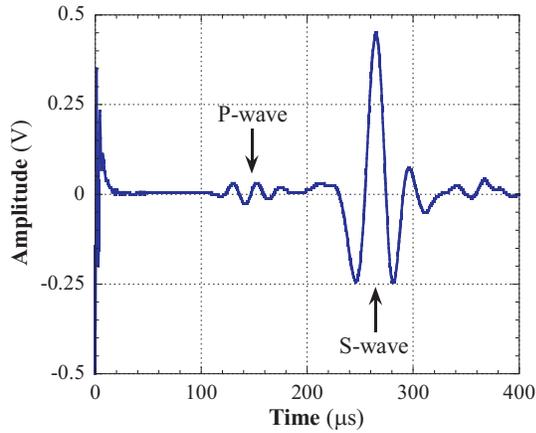}
  \caption{\label{fig:signalUS}Typical ultrasound transmission along the vertical direction ($\sigma_{zz}= $500 kPa),
   excited and dectected by shear transducers.}
\end{figure}

Different transports of sound waves in granular media have been
detailed in the previous work \cite{jia01}. When the wavelength is
much larger than the bead size, the coherent waves propagate as
through an effectively homogeneous medium. Measuring the elastic
wave velocity $V$ allows one to access to the elastic modulus $M$ of
the granular medium by $M = \rho V^{2}$ with $\rho$ the material
density \cite{makse04,jia01,brunet08a}. In our apparatus, the
compressional P-waves and shear S-waves are excited and detected,
respectively, by longitudinal or transversal piezoelectric
transducers of diameter 30 mm. Figure \ref{fig:signalUS} displays a
typical pulsed ultrasound transmission along the vertical direction.
The arrivals of P- and S-wave pulses are well separated, which
allows us to clearly identify the different modes and measure
adequately the wave velocities by the time-of-flight.

\begin{figure}[]
  \includegraphics[width=8cm]{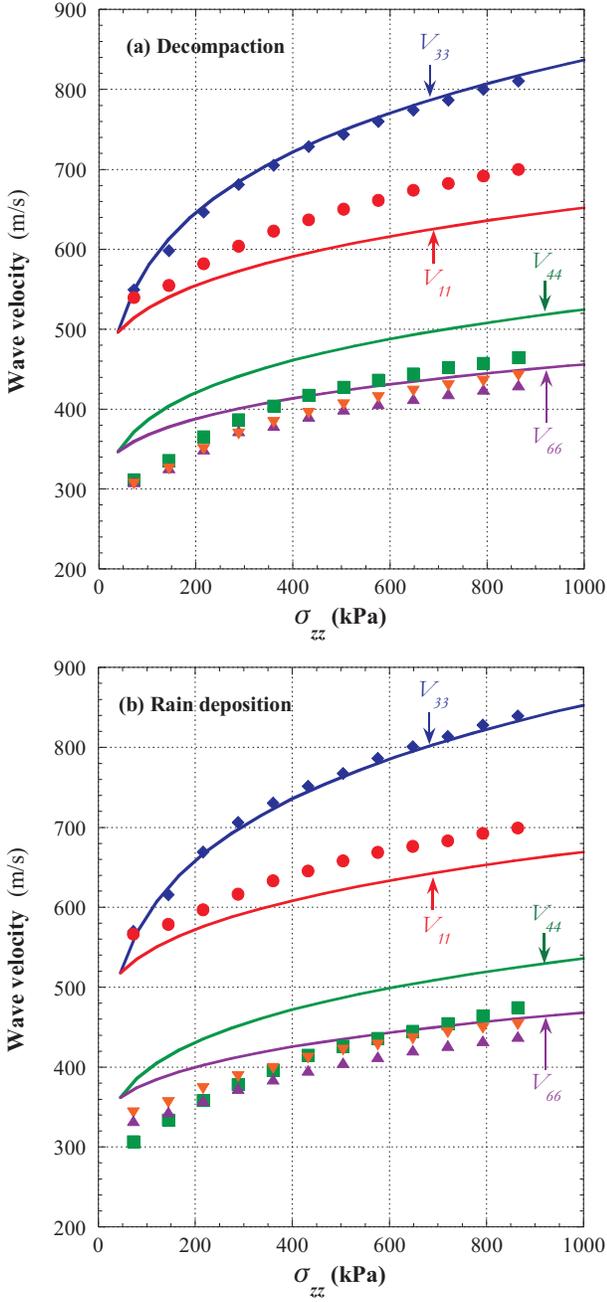}
  \caption{\label{fig:Vii}Wave velocity versus applied stress $\sigma_{zz}$ in the loose pack (a) and
   the dense one (b). Experimental data ($\textcolor{blue}{\Diamondblack : V_{33}}$,
$\textcolor{red}{\bullet :
   V_{11}}$, $\textcolor{Green}{\blacksquare : V_{44}}$, $\textcolor{orange}{\blacktriangledown : V_{55}}$ and
   $\textcolor{violet}{\blacktriangle : V_{66}}$) are compared to the EMT predictions (lines) .}
\end{figure}

For the granular materials under uniaxial loading and the geometry
of the cell considered here, the elastic tensor $C_{IJ}$ relating
the incremental stress $d\sigma_I$ and strain $d\varepsilon_{J}$,
$d\sigma_I = C_{IJ} d\varepsilon_J$, would have a symmetry belonging
to the orthotropic class. In such a case, there are nine independent
constants as shown by the matrix \cite{royer00}:
\begin{equation*}
    \left(
      \begin{array}{c}
        d\sigma_{1} \\
        d\sigma_{2} \\
        d\sigma_{3} \\
        d\sigma_{4} \\
        d\sigma_{5} \\
        d\sigma_{6} \\
      \end{array}
    \right)
    =
    \left(
       \begin{array}{cccccc}
         C_{11} & C_{12} & C_{13} & 0 & 0 & 0 \\
         C_{12} & C_{22} & C_{23} & 0 & 0 & 0 \\
         C_{13} & C_{23} & C_{33} & 0 & 0 & 0 \\
         0      & 0      & 0      & C_{44} & 0 & 0 \\
         0      & 0      & 0      & 0      & C_{55} & 0 \\
         0      & 0      & 0      & 0      & 0       & C_{66} \\
       \end{array}
     \right)
     \left(
       \begin{array}{c}
         d\varepsilon_{1} \\
         d\varepsilon_{2} \\
         d\varepsilon_{3} \\
         d\varepsilon_{4} \\
         d\varepsilon_{5} \\
         d\varepsilon_{6} \\
       \end{array}
     \right).
\end{equation*}
By means of P- and S-waves of different polarization propagating
along the horizontal and vertical directions (i.e. principal axes),
respectively, we can infer the diagonal elastic components ($I = J$)
from the velocity measurement $C_{II} = \rho V_{II}^{2}$. For
determining the off-diagonal components, velocity measurements of
waves propagating along a direction inclined to the principal axes
should be necessary. We display in Fig. \ref{fig:Vii}a and
\ref{fig:Vii}b the wave velocities of several elastic waves measured
a function of the applied stress $\sigma_{zz}$ in our loose pack and
dense one, respectively. Here $V_{11}$ and $V_{33}$ correspond to
the velocities of P-waves propagating along the horizontal and the
vertical ($x$- and $z$-axis) directions. $V_{44}$ denotes to the
velocity of the S-wave propagating vertically, whereas $V_{55}$ and
$V_{66}$ correspond to those of the S-waves propagating horizontally
and being polarized along the vertical and perpendicular ($y$-axis)
directions. Each data presented in Fig. \ref{fig:Vii} results from
an average of six experimental runs and the data dispersion or
error-bar is lower than $2\%$ in the case of the loose pack and
$3\%$ in the dense sample, illustrating thus a good reproducibility
of these acoustic measurements.

As shown in Fig. \ref{fig:Vii}a for the loose pack, there is a clear
difference of P-wave velocity between $V_{11}$ and $V_{33}$ ($>
V_{11}$) going up to $15\%$ when the stress  $\sigma_{zz}$ is
increased to 900 kPa. For the S-wave, the velocity difference
$V_{44} > V_{55}
> V_{66}$  is less important than that of P-wave: indeed, at
$\sigma_{zz} = 900$ kPa $V_{44}$ is about $5\%$ and $8\%$ greater
than $V_{55}$ and $V_{66}$, respectively. In contrast to the
previous works under load superior to a few  MPa
\cite{makse04,johnson98,domenico77}, our acoustic measurements are
realized at much lower stress, ranging from 70 kPa to 900 kPa. In
such range of loading, it is expected that both the solid volume
fraction $\phi$ (Fig. \ref{fig:compac}) and the coordination number
$Z$ would vary little \cite{makse04}. As the de-compaction protocol
tends to create an isotropic fabric in the loose granular sample,
our measurements indicate that the elastic anisotropy observed here
shall stem from the induced stress anisotropy shown in Fig.
\ref{fig:StressRatio}.

For the dense pack prepared by rain deposition, Fig. \ref{fig:Vii}b
displays again a significant difference between the P-wave
velocities $V_{11}$ and $V_{33}$. When the anisotropy of the stress
field is developed, such elastic anisotropy rises to $18\%$ at
$\sigma_{zz} = 900$ kPa. Unlike the P-wave velocity, the S-wave
velocity appears to be sensitive to the fabric anisotropy produced
by rain deposition in the dense granular pack. Contrary to the
measurement in the loose packing, we observe a significant
difference between $V_{44} = 306 \pm 2$ m/s, $V_{55} = 342 \pm 7$
m/s and $V_{66} = 333 \pm 6$ m/s at the lowest stress $\sigma_{zz} =
72$ kPa where the stress field is nearly isotropic (see Fig.
\ref{fig:StressRatio}). Moreover, when $\sigma_{zz}$ is increased
the S-wave velocities evolve differently, but recover a similar
behaviour at high load: a difference of $8\%$ is found again between
$V_{44}$ and $V_{66}$ at $\sigma_{zz} = 900$ kPa. These results
imply a particular sensitivity of S-wave to the fabric anisotropy
which affects the elastic anisotropy together with the stress
anisotropy.
\section{Comparison with the effective medium theory}\label{sec:discussions}
The theory of elasticity a granular pack is primarily based on the
Hertz-Mindlin model of contact between grains \cite{johnson85}. The
macroscopic stress-strain relations are commonly derived using the
effective medium description, where it is assumed that the motion of
grains is affine with the applied macroscopic strain at least on
average, and the distribution of contacts is statistically isotropic
and homogeneous. Because the presence of the tangential forces at
the contacts gives rise to load-displacement relations which are not
only nonlinear but also inelastic, the mechanical response of the
medium, namely stress-strain relations, must be expected to depend
on the entire past history of loading. However, it is shown that the
incremental response of the medium, i.e. the second-order effective
elastic constants and consequently sound velocities are
path-independent of loading \cite{norris97}. Within the framework of
the effective medium approach, Walton \cite{walton87} analyzed the
mechanical responses of a random pack of identical elastic spheres
under isotropic strain and purely uniaxial compression,
respectively. Furthermore the incremental elastic moduli were
derived for these specific initial deformed states. For simplicity
the spheres were assumed to be either \emph{infinitely rough} or
\emph{perfectly smooth} in the calculations; nevertheless these
results provide a physical insight of the effect of friction in real
grain contacts.

As noted above, our measurements were made in the compaction cell
with rigid walls in which the granular sample is subjected neither
to purely isotropic nor uniaxial compression. To compare with the
experimental data obtained in such widely used {\oe}dometric tests,
Johnson \emph{et al.} proposed an analytical model based on the
effective medium theory \cite{johnson98}. This model combines
isotropic and uniaxial strains to describe the stress-induced
elastic anisotropy in transversely isotropic granular materials:
$\varepsilon_{ij} = \varepsilon \delta_{ij} +
\varepsilon_{3}\delta_{i3}\delta_{j3}$  with $\delta_{ij} = 1$ if $i
= j$, otherwise  $\delta_{ij} = 0$. Here the internal strain
$\varepsilon$  stems from the residual stress distribution, namely
an initial isotropic stress state produced by the
preloading-unloading process as mentioned above, whereas the axial
strain $\varepsilon_3$ results from the applied stress
$\sigma_{zz}$. As $V_{55}$ differs from $V_{44}$ only by $5\%$ (Fig.
\ref{fig:Vii}), the approximation of the transverse isotropy used in
this model would be adequate for interpreting our measurements. The
predictions of such a model reduce to the results for purely
isotropic compression in the limit $\varepsilon_3 \rightarrow 0$ and
to those for purely uniaxial compression in the opposite limit
$\varepsilon \rightarrow 0$, respectively \cite{walton87}.

Fig. \ref{fig:StressRatio} presents the ratio $\sigma_{zz} /
\sigma_{xx}$ computed for the packs of both bonded ("rough") and
frictionless ("smooth") spheres, as a function of the applied stress
$\sigma_{zz}$. The latter case is obtained by canceling the
tangential contact stiffness in the model, i.e. $C_t \rightarrow 0$.
As described in eqs. 21 and 22 of \cite{johnson98}, the stress
tensor $\sigma_{ij}$ is a function of several parameters,
$\sigma_{ij} = \sigma_{ij} (\varepsilon , \varepsilon_3, Z, \phi ,
\mu_g, \nu_g)$ where $Z$ is the coordination number, $\mu_g$ (=
24MPa) is the shear modulus and $\nu_g$ (= 0.2) is the Poisson ratio
of the glass bead. For a given $C_t$ the ratio $\sigma_{zz} /
\sigma_{xx}$ versus the applied stress $\sigma_{zz}$ is only
parameterized by the glass bead property $\mu_g$ and $\nu_g$ (Fig.
\ref{fig:StressRatio}). The axial strain $\varepsilon_3$ is measured
from the displacement of the top piston versus $\sigma_{zz}$ and the
residual isotropic strain $\varepsilon$ is deduced from the
isotropic residual stress 70 kPa measured at $\varepsilon_3 = 0$,
yielding $\varepsilon = 1.5\times10^{-4}$.

We observe from Fig. \ref{fig:StressRatio}, that the theoretical
curve calculated with the frictionless spheres ($C_t \rightarrow 0$)
is faithful to the qualitative trend observed in our experiments and
is overall in agreement with the measured data; at the high load the
stress ratio approaches to the limit value of uniaxial loading, i.e.
$\sigma_{zz} / \sigma_{xx} = 4$ \cite{walton87,johnson98}. In
contrast, Fig. \ref{fig:StressRatio} and the inset show that the
no-slip assumption with bonded spheres gives rise to a huge
overestimation of this stress ratio, indicating the crucial role of
the contact interaction law in the effective medium model for
describing adequately the evolution of the stress field anisotropy.
We may understand the discrepancy found with the bonded spheres by
the following picture. At large deformation applied here of the
order of $10^{-3}$, the grains need to rearrange in order to relax
the tangential stress accumulated at the contacts. The no-slip
assumption with infinitely frictional spheres forbidden such a
process and hence fail, whereas the effective medium analysis with
frictionless spheres allow the stress relaxation and provide thus an
adequate description of the experiments.

Let us now examine the applicability of the effective medium
approach to the wave velocity measurements. Given a stress state
$\sigma$ and corresponding macroscopic deformation $\varepsilon$,
the elastic moduli can be derived from the incremental stress of the
medium subject to an incremental strain by $d\sigma_{ij} = C_{ijkl}
d\varepsilon_{kl}$. Following the work of \cite{johnson98}, yields
\begin{eqnarray}
C_{ijkl}&=&\frac{3\phi Z}{4\pi^{2}R^{1/2}B(2B+C)} \langle \xi^{1/2}
\{2Cn_{i}n_{j}n_{k}n_{l}\nonumber\\*
        & &+B\left(\delta_{ik}n_{j}n_{l} + \delta_{il}n_{j}n_{k} +
\delta_{jl}n_{i}n_{k} + \delta_{jk}n_{i}n_{l}\right)\} \rangle
\label{eq:Cij}
\end{eqnarray}
where $B=\frac{1-\nu_{s}}{2\pi \mu_{s}} \: , \:
C=\frac{\nu_{s}}{2\pi\mu_{s}} \: , \:
\xi=-\mathbf{n}\cdot\boldsymbol{\epsilon}\cdot\mathbf{n}R$, and
$\delta_{ij}$ denotes the Kronecker symbol. The brackets $\langle
\rangle$ represent an average over all unit vector $\mathbf{n}$
uniformly distributed since the distribution of contact angle (i.e.
granular fabric) is assumed to be isotropic. To determine the wave
velocities from the elastic moduli in eq. \ref{eq:Cij}, we perform
the calculations by two steps. Firstly, we compute the stress state
of a frictionless spheres packing ($C_t = 0$) created by the large
axial deformation, which provides an adequate description of the
experimental stress field shown in Fig. \ref{fig:StressRatio}. Then,
we consider the appropriate value of $C_t$ ($\neq 0$) according the
Hert-Mindlin theory to calculate the elastic moduli by eq.
\ref{eq:Cij}. Note that this procedure of computation is also used
in many numerical simulations where the friction between the
particles is turned off during the packing preparation and is turned
on to measure the incremental response \cite{makse04}. Fig.
\ref{fig:Vii}a and \ref{fig:Vii}b illustrate the calculated wave
velocities in both the loose packing and the dense one. The
agreement between the theoretical predictions and experimental
results remains fairly good (better than $10\%$), for the present
effective medium model employing only one adjustable parameter, i.e.
$Z = 6$ in the dense pack and $Z = 5.5$ in the loose pack.
\section{Discussions}\label{sec:discussions}
\subsection{Stress dependence of P-wave velocities}
Based on the Hertz-Mindlin theory of contact and the assumption of
isotropic fabric in the sphere pack, the effective medium theory
(EMT) predicts a power-law dependence of the elastic modulus $M \sim
P^{1/3}$ (or $V \sim P^{1/6}$) on isotropic compression $P$
\cite{goddard90,makse04,duffy57,digby81}. The same power law scaling
is found for the pure uniaxial loading (i.e. $\varepsilon_{ij} =
\varepsilon_3 \delta_{i3} \delta_{j3}$) as $M \sim
\sigma_{zz}^{1/3}$ or $M \sim \sigma_{xx}^{1/3}$ since the stress
ratio $\sigma_{zz} / \sigma_{xx}$ is constant \cite{walton87}.
Numerous acoustic velocity measurements in sands and glass bead
packs showed however that the exponent in the power-law scaling may
not be constant, varying from $1/4$ at low pressure to $1/6$ at high
pressure \cite{goddard90,jia99,richart70,gilles03}. Several
mechanisms have been proposed to explain this discrepancy, including
the contact recruitment by buckling of particle chains or the
conical contact due to irregular surfaces \cite{goddard90}, the
effect of soft shell on coated spheres \cite{degennes96} and the
fluctuation of the stress field \cite{velicky02}. However, for the
\emph{preloaded} glass bead packs and the range of the applied
stress considered here, the contact recruitment is expected to have
a negligible effect, as that suggested by the above density
measurement (Fig. \ref{fig:compac}).

\begin{figure}[]
  \includegraphics[width=8cm]{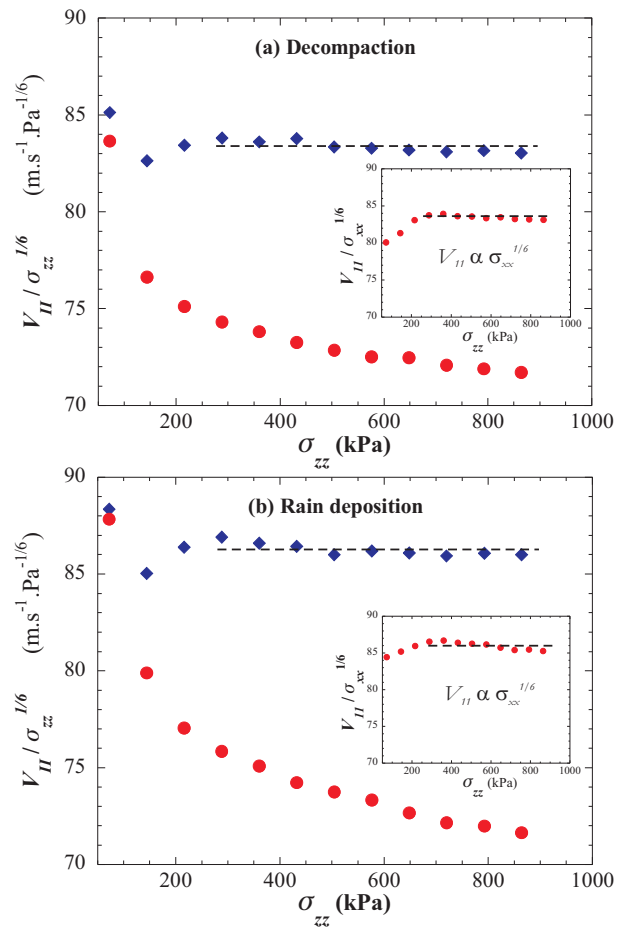}
  \caption{\label{fig:P-Scaling}Stress dependence of P-wave
  velocities propagating vertically $\left ( \textcolor{blue}{\Diamondblack : V_{33}}\right )$
  and horizontally $\left ( \textcolor{red}{\bullet : V_{11}} \right )$ in the loose pack (a) and the dense one (b).
   The dashed lines correspond to the predictions by Hertz-Mindlin theory of contact.}
\end{figure}

We investigate here the influence of the stress anisotropy on the
scaling behaviour of the P-wave velocity. To do this, we displays in
Fig. \ref{fig:P-Scaling} the rescaled P-wave velocities $V_{33}/
\sigma_{zz}^{1/6}$ and $V_{11}/\sigma_{zz}^{1/6}$ as a function of
the applied axial stress $\sigma_{zz}$, obtained in the loose and
dense packs, respectively. For $\sigma_{zz} > 300$ kPa, $V_{33}$ is
seen to follows adequately the scaling of $\sigma_{zz}^{1/6}$, as
expected within the framework of the Hertz-Mindlin theory. By
contrast, $V_{11}$ deviates drastically from the $\sigma_{zz}^{1/6}$
scaling, which is due to the evolution of stress-field anisotropy
illustrated in Fig. \ref{fig:StressRatio}. Indeed, if rescaling
$V_{11}$ with the stress $\sigma_{xx}$ parallel to the propagation
direction, we recover a scaling behaviour $V_{11} \sim
\sigma_{xx}^{1/6}$ (insets to Figs. \ref{fig:P-Scaling}a and
\ref{fig:P-Scaling}b), similar to those for $V_{33} \sim
\sigma_{zz}^{1/6}$ in both the loose and the dense packs having the
different granular fabric.
\subsection{Stress and fabric dependences of S-wave velocities}
\begin{figure}[]
  \includegraphics[width=8cm]{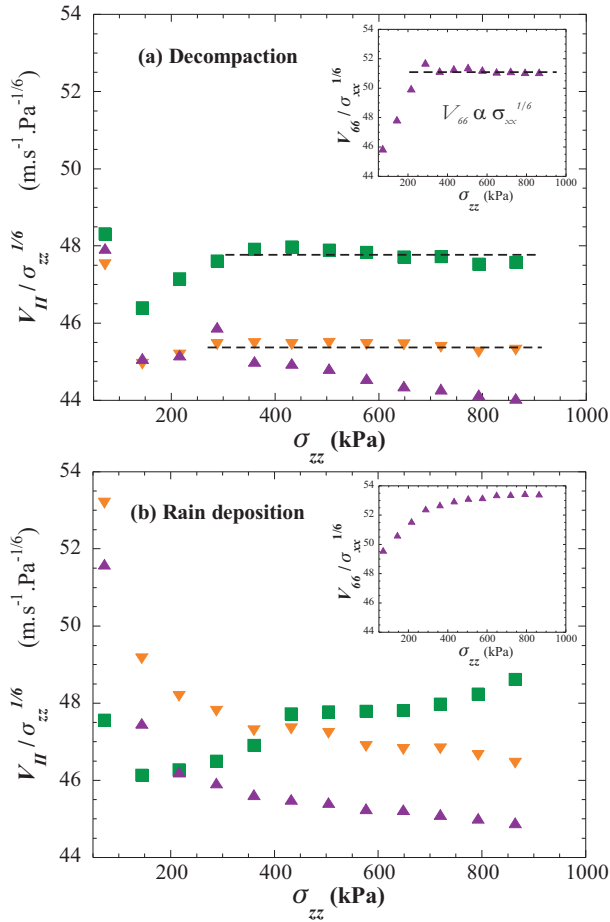}
  \caption{\label{fig:S-Scaling}Stress dependence of S-wave
  velocities propagating vertically $\left ( \textcolor{Green}{\blacksquare : V_{44}}\right )$
  and horizontally $\left ( \textcolor{orange}{\blacktriangledown :
V_{55}}
   \textrm{ and } \textcolor{violet}{\blacktriangle : V_{66}} \right )$ in the loose pack (a) and the dense one (b).
   The dashed lines correspond to the predictions by the Hertz-Mindlin contact theory.}
\end{figure}

Compared to the P-wave velocity, the stress dependence of the S-wave
velocity is more complicated and strongly influenced by the fabric
of the pack.  As mentioned above (in sec. \ref{sec:experiment}),
S-waves are not only sensitive to the stress field anisotropy but
also to the fabric anisotropy. Let us first examine the scaling
behaviour of the shear velocities versus $\sigma_{zz}$ in the loose
bead pack with isotropic fabric. As shown in Fig.
\ref{fig:S-Scaling}a, both $V_{44}$ and $V_{55}$ scale roughly as a
power law i.e. $\sim \sigma_{zz}^{1/6}$ for $\sigma_{zz} > 300$ kPa,
which agree fairly well with the stress-dependence predicted by the
Hertz-Mindlin theory of contact. Compared to the P-wave velocity
$V_{11}$, the scaling behaviours of the shear velocity $V_{55}$ is
somewhat surprising, however, it may be understood by a heuristic
picture which accounts for both the propagation direction and the
polarization. Indeed, unlike the P-wave propagating in the
horizontal direction ($V_{11}$), the S-wave travelling along the
$x$-axis is polarized vertically along the $z$-axis ($V_{55}$) and
it would also be affected by the stress component $\sigma_{zz}$.
This scenario is consistent with the preceding observation showing
that the S-wave velocity is less sensitive to the stress-field
anisotropy than the P-wave velocity is (shown in Fig.
\ref{fig:Vii}). However, for the S-wave travelling along the
$x$-axis but polarized horizontally along the $y$-axis ($V_{66}$),
Fig. \ref{fig:S-Scaling}a shows a scaling behaviour of $V_{66}$
deviated from $\sigma_{zz}^{1/6}$ but being consistent with the
power law $V_{66} \sim \sigma_{xx}^{1/6}$ (inset to Fig.
\ref{fig:S-Scaling}a), a situation reminiscent to that of the P-wave
($V_{11}$).

We now turn on the scaling behaviour of the S-wave velocities in the
dense pack presenting the fabric anisotropy obtained by rain
deposition. Contrary to the behaviours observed in the loose packing
with isotropic fabric, there is no clear stress dependence of the
shear velocities $V_{44}$, $V_{55}$, and $V_{66}$ in the range of
the applied stress (Fig. \ref{fig:S-Scaling}b and the inset). This
observation reveals the extreme sensitivity of the shear wave
velocity to the fabric of the granular pack. However, at the present
stage, it is difficult to speculate the scaling behaviour of the
S-wave velocity on stress, due to the interplay between the
evolution of the stress anisotropy and the inherent fabric
anisotropy. Indeed, the latter parameter is not included in most of
analytical models within the framework of the effective medium
theory.
\subsection{Correlation between induced elastic anisotropy and stress-field anisotropy}
In the above sections, we have observed that the elastic moduli and
consequently sound velocities (P- and S-waves) depend on both the
induced stress-field anisotropy and the inherent fabric of the
granular sample. The protocols of the sample preparation used here,
i.e. de-compaction and rain deposition, allow us to obtain two
distinct granular fabrics which also differ in the packing density
$\phi$ as shown in Fig. \ref{fig:compac} and probably in the
coordination number $Z$. As suggested by eq. (\ref{eq:Cij}), we may
investigate the correlation between the induced stress anisotropy
and the elastic anisotropy via the ratio of elastic components
dropping thus the parameters $\phi$ and $Z$.

\begin{figure}[b]
  \includegraphics[width=8cm]{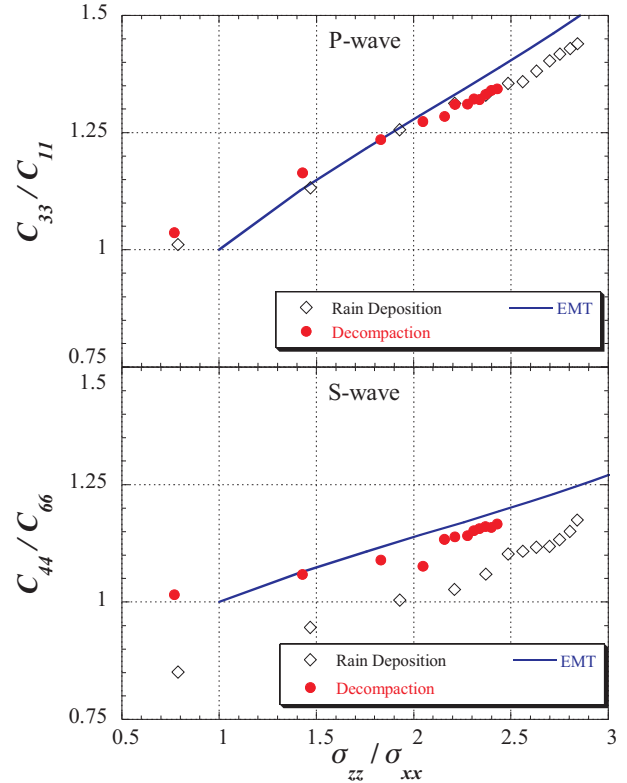}
  \caption{\label{fig:AniCorr}Correlation between the induced elastic anisotropy
   and stress-field anisotropy in the granular packs of different fabric.}
\end{figure}

Fig. \ref{fig:AniCorr}a displays the ratio of the compressional
moduli $C_{33} / C_{11}$ versus the induced stress anisotropy
$\sigma_{zz} / \sigma_{xx}$ for two granular samples with different
fabric (loose and dense). The experimental data agree well with the
effective medium model (eq. \ref{eq:Cij}), demonstrating that the
anisotropy of compressional moduli is principally determined by the
stress field anisotropy and is much less sensitive to the fabric of
the granular pack. Also, we depict the evolution of the shear
modulus ratio $C_{44} / C_{66}$ as a function of $\sigma_{zz} /
\sigma_{xx}$ in Fig. \ref{fig:AniCorr}b. These results confirm again
the correlation of the elastic anisotropy with the stress-field
anisotropy, though the anisotropy is less pronounced for the shear
modulus than for the compressional one. In contrast to the latter,
the shear moduli are however very sensitive to the granular fabric.
For the loose pack with isotropic fabric, the ratio of the shear
moduli is in good agreement with the effective medium theory model
based on the assumption of the isotropic distribution of contact
angle.
\section{Conclusion}\label{sec:conclusion}
In summary, we have studied the anisotropic elasticity of dry glass
bead packs using ultrasonic measurements. Both the influences of the
fabric anisotropy and the stress anisotropy are investigated thanks
to the granular samples of different inherent fabric under uniaxial
loading in an {\oe}dometric cell. The stress anisotropy is
characterized from the horizontal and vertical force measurements,
whereas the elastic moduli are determined from velocity measurements
of P- and S-waves propagating along the vertical and horizontal
directions, respectively. For the range of the applied stress in
this work, our results show clearly that the compressional modulus
anisotropy is more sensitive to the stress field anisotropy while
the shear modulus anisotropy is more sensitive to the fabric
anisotropy.

We have tested the applicability of the effective medium theory to
sound propagation associated with the incremental deformation. The
predictions by the analytical model based on the affine
approximation agree well with the measurements of sound velocity.
Moreover, our velocity measurements of compressional waves confirm
the scaling behaviour on stress, $V_P \sim P^{1/6}$, where $P$ is
the normal component of the stress along the propagation direction.
As for the shear waves, the stress dependences of velocities are
more complex, being also strongly influenced by the fabric
anisotropy which is not considered in existing analytical EMT
models.

The present effective medium model is also applied to analyze the
finite deformation of the granular pack. Our stress-strain
measurements show the breakdown of this model to describe adequately
the deformed state at large deformation. However, such discrepancy
between theory and experiment can significantly be reduced when the
friction between grains is switched off. As indicated previously
\cite{makse04,jenkins04}, our observation confirms that the model
based on the affine approximation fails because it prevents the
relaxation of the tangential stress at the grain contact, especially
for large deformation. Other effective medium models are required
for describing the large deformations of granular packs
\cite{jiang07,hicher06}.

\newpage 

\end{document}